\documentclass[final,3p,times]{elsarticle}

\usepackage{amsmath,amssymb,amsfonts}
\usepackage{algorithmic}
\usepackage{graphicx}
\usepackage{textcomp}
\usepackage[table,xcdraw,dvipsnames]{xcolor}
\usepackage{amsmath}
\usepackage{siunitx}
\usepackage{orcidlink}
\usepackage{hyperref}
\usepackage[capitalise]{cleveref}
\usepackage{subfigure}
\usepackage{booktabs}
\usepackage{upgreek}
\usepackage{bm}

\hypersetup{hidelinks}

\begin{document}

\begin{frontmatter}

\title{Design and Simulation of a 4H-SiC Low Gain Avalanche Diode with Trench-Isolation}

\author[inst1]{Sebastian Onder\corref{cor1}}
\ead{sebastian.onder@oeaw.ac.at}
\author[inst1]{Philipp Gaggl\corref{cor1}}
\ead{philipp.gaggl@oeaw.ac.at}
\author[inst1]{J{\"u}rgen Burin}
\author[inst1]{Andreas Gsponer}
\author[inst2]{Matthias Knopf}
\author[inst1]{Simon Waid}
\author[inst3]{Neil Moffat}
\author[inst3]{Giulio Pellegrini}
\author[inst1]{Thomas Bergauer}

\affiliation[inst1]{organization={Institute for High Energy Physics, Austrian Academy of Sciences},
            addressline={Nikolsdorfer Gasse 18}, 
            postcode={1050}, 
            city={Vienna},
            country={Austria}}
\affiliation[inst2]{organization={Atominstitut, TU Wien},
            addressline={Stadionallee 2}, 
            postcode={1020}, 
            city={Vienna},
            country={Austria}}

\affiliation[inst3]{organization={Instituto de Microelectronica de Barcelona, IMB-CNM-CSIC},
            postcode={08193}, 
            city={Barcelona},
            country={Spain}}

\cortext[cor1]{corresponding author}

\begin{abstract}
We present the design and simulation of a $\SI{30}{\micro\meter}$ thick 4H-SiC Low Gain Avalanche Diode (LGAD) optimized for high-voltage operation. A $\SI{2.4}{\micro\meter}$ thick epitaxially grown gain layer enables controlled internal amplification up to $\SI{1}{\kilo \volt}$ reverse bias, while maintaining full depletion below $\SI{500}{\volt}$. Electrical characteristics, including I-V, C-V, and gain behavior, were simulated in Synopsys Sentaurus Technology Computer-Aided Design (TCAD) using a quasi-1D geometry and verified across process-related variations in gain layer parameters. To ensure high-voltage stability and proper edge termination, a guard structure combining deep etched trenches and deep $p^+$ junction termination extension (JTE) implants was designed. TCAD simulations varying the guard structure dimensions yielded an optimized design with a breakdown voltage above $\SI{2.4}{\kilo \volt}$. A corresponding wafer run is currently processed at IMB-CNM, Barcelona.
\end{abstract}


\end{frontmatter}

\section{Introduction}

The growing demand for silicon carbide (SiC) devices in the power electronics and automotive industry has improved both the material’s availability and fabrication techniques. This progress has renewed interest in using SiC, especially its prominent polytype 4H-SiC, as a base material for radiation sensors \cite{nava2008, denapoli2022, capan2022b}. Due to its significantly larger bandgap compared to the commonly used silicon (Si), 4H-SiC exhibits low dark current levels ($<\SI{10}{\pico\ampere}$), even after irradiation and at room temperature \cite{Gaggl2022, Gsponer_2023}. Its high breakdown field and charge carrier saturation velocity enable fast signal generation and operation at considerably high reverse bias, which could be leveraged for fast timing detectors in high-energy physics (HEP).

However, current epitaxial growth techniques commonly used within industry yield relatively high doping concentrations ($\geq \SI{e14}{\centi \meter^{-3}}$). The subsequently large depletion voltages render thicker active regions impractical due to challenges regarding high-voltage stability. This currently limits SiC particle detectors to active thicknesses of around $\SI{100}{\micro\meter}$ \cite{denapoli2022}. Furthermore, 4H-SiC has a higher electron-hole pair creation energy than Si (Si: $\SI{3.6}{\electronvolt}$, 4H-SiC: $\sim \SI{7.8}{\electronvolt}$). A minimum ionizing particle (MIP) only generates about 57 electron-hole pairs per $\si{\micro\meter}$ in SiC, roughly two-thirds relative to Si \cite{paper_HEPHY_0}. Conventional SiC pn-junction sensors thus only yield small signal amplitudes that are insufficient to detect MIPs, gravely limiting the attractiveness of the material for the HEP community. 

In contrast, Si-based low gain avalanche diodes (LGADs) have been successfully used in HEP experiments as fast timing detectors by taking advantage of an additional highly doped gain layer, enabling internal signal amplification due to charge multiplication \cite{Pellegrini_2014}. Successfully realizing such a design using 4H-SiC potentially compensates for the inherently small signals in conventional SiC sensors. Considering its naturally fast signal generation and a low dark current, 4H-SiC is an ideal candidate for LGAD development and detectors with ultra-fast timing performance. Several groups have already demonstrated prototype SiC-LGADs. The SICAR project designed devices using a bevel-edge process for isolation, reporting a gain of 2-3 \cite{Zhao_2024}. A production from LBNL and NCSU used a similar isolation approach, reporting a gain of 7-8 and time resolution of $<\SI{35}{\pico \second}$ based on UV-TCT measurements \cite{Yang_2025}. The latest production of 4H-SiC LGADs from FNSPE CTU and FZU CAS in collaboration with ON Semiconductor \cite{Novotny2025} used implanted junction termination extensions (JTEs) for isolation and managed to reach an epitaxial layer doping down to $\SI{5e13}{\centi \meter^{-3}}$. They reported a gain of 10-100.

We present a TCAD-based design of a 4H-SiC LGAD, featuring trench isolation combined with deep-implanted JTEs optimized for high-voltage operation. A corresponding wafer run has already been fabricated and received, while several processing steps, such as metallization and trench etching, remain pending at IMB-CNM in Barcelona, Spain \cite{URL_CNM}.

\section{Device Design}
\label{sec:device_design}

\subsection{Epitaxial Structure}
\label{sec:epi_structure}

\begin{figure}
    \centering
    \includegraphics[width=0.65\linewidth]{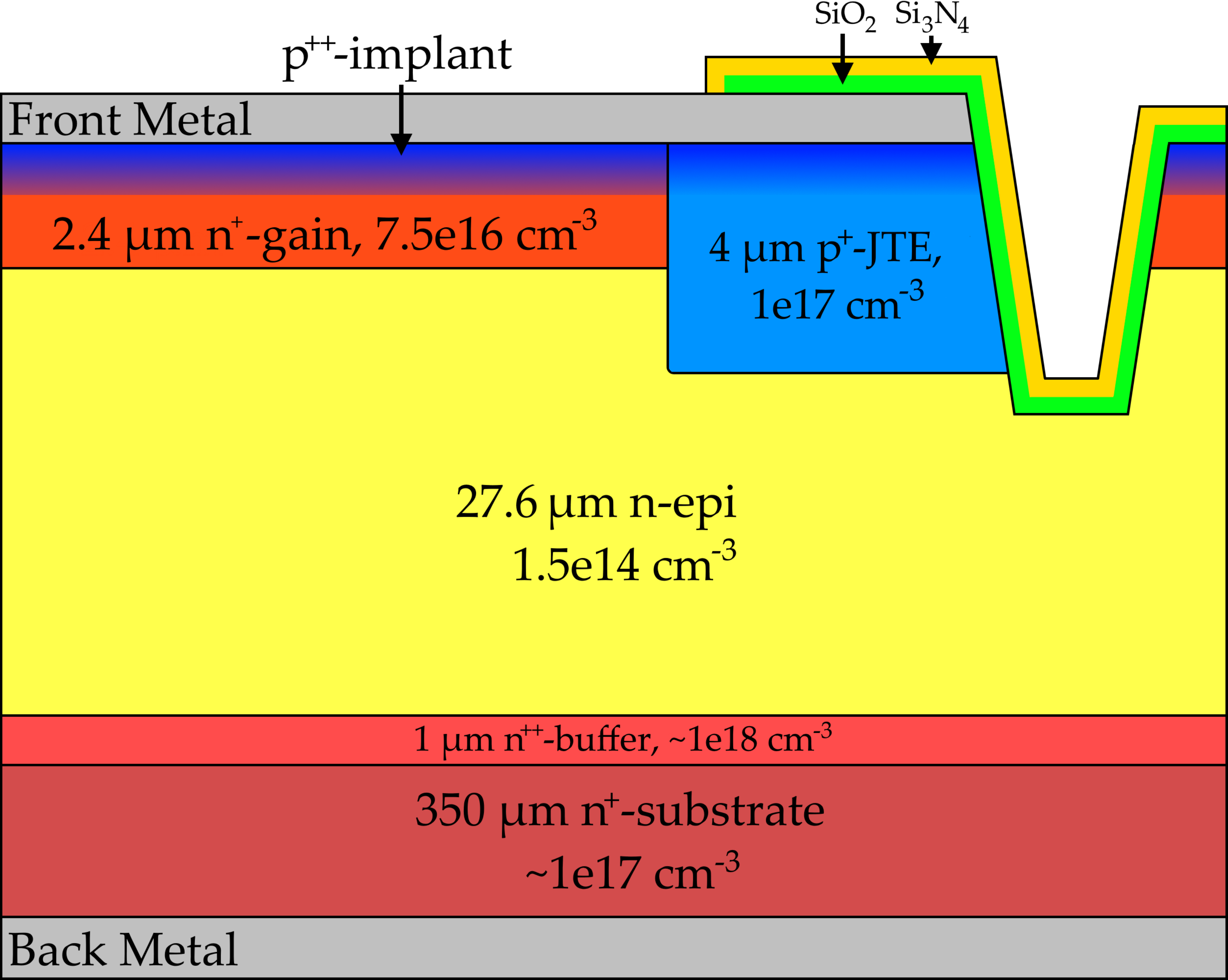}
    \caption{Cross-section of the epitaxial structure of the 4H-SiC LGAD design (not true to scale). The edge of the front contact and the subsequent guard structure are shown. The latter consists of an etched trench and a JTE implant to further distribute fringe fields.}
    \label{fig:epi_structure}
\end{figure}

Fig.\ref{fig:epi_structure} shows a cross-section of the epitaxial structure for the proposed 4H-SiC LGAD design. The device is grown on a low-quality $n^{+}$ substrate with a doping concentration of approximately $\SI{e17}{\centi \meter^{-3}}$. On top, a $\SI{1}{\micro\meter}$ thick, highly doped ($\sim\SI{e18}{\centi\meter^{-3}}$), $n^{++}$ buffer layer serves as a field stop. The active region consists of a highly resistive, $\SI{27.6}{\micro \meter}$ thick, $n$-type epitaxial layer with a doping concentration of $\SI{1.5e14}{\centi\meter^{-3}}$, followed by a $\SI{2.4}{\micro\meter}$ $n^{+}$ gain layer with a doping of $\SI{7.5e16}{\centi\meter^{-3}}$. Finally, a $p^{++}$ implant establishes the pn-junction, while simultaneously providing an ohmic contact to the front metallization.

As the excessively high implantation energies required to implant sufficiently deep ($\approx \SI{1}{\micro\metre}$) were not available to us, an epitaxial growth step was utilized to deposit the gain layer across the entire wafer. 
To that end, a non-buried approach was chosen to eliminate the potential risk of residual doping of unknown intensity subsequent to the gain layer. The above-reported gain layer thickness thus spans from the very top of the device, even though parts of it are compensated by the  $p^{++}$-implant. The doping concentration of the n-epi layer represents the lowest possible concentration the manufacturer could reliably provide, aiming to minimize the full depletion voltage. The device thickness, as well as the thickness and doping concentration of the gain layer, were optimized through extensive parameter sweeps in Synopsys Sentaurus TCAD \cite{URL_Sentaurus}. The target requirements included a full depletion voltage below $\SI{500}{\volt}$, stable operation up to $\SI{1}{\kilo \volt}$ reverse bias, and a signal gain (when compared to a common PIN diode of equal thickness) between 2 and 10.

Ten wafers with the finalized structure have been manufactured and received. According to the manufacturer, the gain layer has a thickness tolerance of $\SI{0.2}{\micro \meter}$ and a doping variation of up to $\SI{10}{\percent}$, which justifies the decision to utilize a relatively thick gain layer with moderate doping, ensuring robustness against such uncertainties.

\subsection{Guard Design}
\label{sec:guard_design}

Because the gain layer is grown homogeneously across the full 4H-SiC wafer and has a relatively large thickness, conventional guard structures using shallow $p$-implants are insufficient to isolate and shield individual LGADs properly. To ensure reliable edge termination, the design incorporates etched trenches surpassing the gain layer thickness, as shown in Fig.\ref{fig:epi_structure}. These trenches, soon to be processed at IMB-CNM, are unfilled and passivated along the walls via deposition of a $\mathrm{SiO_2}$/$\mathrm{Si_3N_4}$ stack. The trenches enclose the entire active area of the circular diodes with a diameter of $\SI{500}{\micro\meter}$.

Initial simulations of the trench geometry, including trench width, depth, angle, and potential etching edges, revealed the trench angle to play a critical role in breakdown behavior. Specifically, trench wall angles exceeding $\SI{90}{\degree}$ (relative to the wafer surface) accumulate a high electric field at the inner trench corner located on the surface of the substrate, provoking premature breakdown below $\SI{1}{\kilo \volt}$ reverse bias. To mitigate this, an additional $p^+$ JTE was introduced at the inner side of the trench. Early simulations incorporating shallow JTE implants (not surpassing the gain layer), as fabricable at IMB-CNM, already showed notable improvements, with breakdown voltages rising above $\SI{1}{\kilo \volt}$. However, through a collaboration with mi2-factory \cite{URL_mi2}, sufficiently high implantation energies to penetrate the gain layer (up to $\SI{4}{\micro \meter}$) became accessible. Introducing such deep JTEs with an implant depth of $\SI{4}{\micro \meter}$ and a uniform doping concentration around $\SI{e17}{\centi \meter^{-3}}$ into the model significantly improved breakdown stability by relocating electric field peaks into lower doped regions. Such a design also covers potential uncertainties in manufacturing, as simulations only suggest a slight dependence of the breakdown voltage on trench angle and potential etching edges. Guard structure design parameters were optimized via TCAD breakdown simulations, details of which are presented in section \ref{sec:breakdown_results}. The final design parameters are summarized in Tab.\ref{tab:guard_design_parameters}.

\begin{table}[ht]
\centering
\begin{tabular}{r|ccc}
 & width (\si{\micro\metre}) & depth (\si{\micro\metre}) & doping (\si{\per\centi\metre\cubed}) \\ \hline
trench & 5/10/15* & 7 & - \\
JTE & 30 & 4 & 10\textsuperscript{17} \\
\end{tabular}
\caption{Optimized guard structure design parameters according to performed TCAD breakdown simulations. *Value confined by the manufacturer, despite simulations suggesting better performance at larger widths (see section \ref{sec:breakdown_results}).}
\label{tab:guard_design_parameters}
\end{table} 

\section{Simulation}
\label{sec:simulation}

The simulation results presented in the following were obtained using Sentaurus TCAD SDevice. We evaluated the current (I-V), capacitance (C-V), and signal gain characteristics as a function of the applied reverse bias to reach the LGAD design presented in section \ref{sec:epi_structure}. These simulations, as described in section \ref{sec:sim_wo_guarding}, contained a wide range of design parameters and considered manufacturing tolerances as given by the vendor. Any fringe structures, such as the guarding, however, were omitted due to their negligible effect on the relevant simulation output. After finalizing the LGAD design, additional simulations incorporating a variety of guard structures were carried out (section \ref{sec:sim_w_guarding}). In all cases, the gain layer was implemented via a box profile and according to specifications given in section \ref{sec:epi_structure}. Any simulated metallization used titanium as electrode material. 

\subsection{Simulation Setup and Method}
\label{sec:sim_setup}

\subsubsection{Gain Layer Design}
\label{sec:sim_wo_guarding}

To reduce computation time, the simulations of I-V, C-V, and gain characteristics utilized a quasi-1D geometry of \SI{1}{\micro\metre} width, omitting any edge structures. Apart from the low-quality substrate, the full vertical epitaxial stack shown in Fig.\ref{fig:epi_structure} was considered. The low structure complexity allowed for extensive meshing along the vertical (depth) axis to accurately resolve doping profiles and subsequently emerging potentials. Laterally, four grid lines were used in order to avoid encountered bugs with the \textit{HeavyIon} model, utilized to model transient signal response to particle hits. To determine the signal gain, a conventional PIN diode identical to the LGAD model, but omitting the gain layer, was simulated to serve as reference.

I-V and C-V characteristics were obtained simultaneously using quasistationary voltage ramps up to $\SI{1}{\kilo \volt}$ reverse bias, and contained a termination option in case current levels exceeded device breakdown. Capacitance behavior was determined via a \textit{Mixed-Mode} AC simulation at $\SI{10}{\kilo \hertz}$. The resulting potentials and electric fields were saved every $\SI{10}{\volt}$, and later used as input for transient signal simulations. To evaluate the signal response to a particle hit, transient simulations were performed using the \textit{HeavyIon} model, emulating a MIP. A linear energy transfer (LET) factor of $\SI{9.15}{\pico\coulomb\per\micro\meter}$, gaussian lateral width of \SI{0.15}{\micro\metre}, and temporal width of $\SI{0.5}{\pico\second}$ were chosen within the \textit{HeavyIon} parameter set, to reproduce a realistic charge deposition of 57 electron-hole pairs per $\si{\micro \meter}$, the most probable value for 4H-SiC \cite{paper_HEPHY_0}. The signal gain was then calculated by normalizing the integrated current signal from the respective LGAD structure to that of the reference PIN diode.

Due to the low intrinsic carrier concentration in 4H-SiC, all simulations required adapted convergence, error, and solver settings to prevent convergence issues \cite{URL_SiC_Sim}. An artificial carrier generation rate using the \textit{ConstantCarrierGeneration} statement was applied to emulate an electronic noise baseline of approximately $\SI{1}{\pico \ampere}$, consistent with measured data from reference samples \cite{Gaggl2022, Gsponer_2023}. The \textit{Okuto} impact ionization parameter set was used to model charge multiplication. Anisotropy was considered using a conventional cutoff angle of 4$^{\circ}$. For further details on the physical models and custom 4H-SiC material parameters used in the simulations, as well as tightened error and convergence criteria, see \cite{URL_SiC_Sim, URL_LGAD}.

\subsubsection{Guard design}
\label{sec:sim_w_guarding}

The breakdown behavior of the finalized LGAD design was investigated via 2D simulations, incorporating the full vertical epitaxial layer stack as described previously. To reduce computation time, the simulated geometry was limited to the immediate vicinity of the guard structures, i.e., the JTE implant and the trench. As such, only a small part ($\SI{15}{\micro\metre}$) of the pad contact was included in the simulation. The distance between the outer boundary and the end of the guard structure was chosen such that no residual electric field was present at the boundary during breakdown, allowing for full propagation of all emerging fringe fields. The vertical doping profile for a $\SI{4}{\micro \meter}$ deep JTE implant was provided by mi2-factory. The passivation was modeled as a $\SI{700}{\nano\metre}$ thick $\mathrm{SiO_2}$, and a $\SI{500}{\nano\metre}$ thick $\mathrm{Si_3N_4}$ layer, consistent with manufacturing specifications. Passivation openings at the pad contact are according to the design rules given by IMB-CNM.

Preliminary simulations indicated a trench wall angle deviating from $\SI{90}{\degree}$ to have negligible influence on the breakdown voltage, as long as the JTE implants surpassed the gain layer. Therefore, a constant sidewall angle of $\SI{96}{\degree}$, taken from preliminary etching tests on 4H-SiC bulk wafers by IMB-CNM, was assumed for all subsequent simulations. The meshing procedure was adapted to the local complexity of the structure, yielding a moderately fine mesh within the gain and passivation layers, and a very fine mesh at critical regions such as material interfaces, doping transitions, and corners of the trench and JTE implant.

Breakdown voltages were extracted using quasistationary simulations with a reverse bias ramp. A current value that could reliably be attributed to device breakdown was obtained through test simulations and used as a termination trigger for all simulations. The breakdown voltage was then determined as the last voltage simulated at termination. A parametric sweep of JTE width, trench width, and trench depth was performed to identify the most stable configuration under high-voltage operation. All material parameters, physical models, and error and convergence criteria were kept identical to the simulations described in section \ref{sec:sim_wo_guarding}. For the passivation layers, default material parameters provided by Sentaurus TCAD were used.

\subsection{I-V, C-V and Gain Characteristics}
\label{sec:I-V_C-V_Gain}

\begin{figure}[ht!]
    \centering
    \subfigure[I-V]{\includegraphics[width=0.7\linewidth]{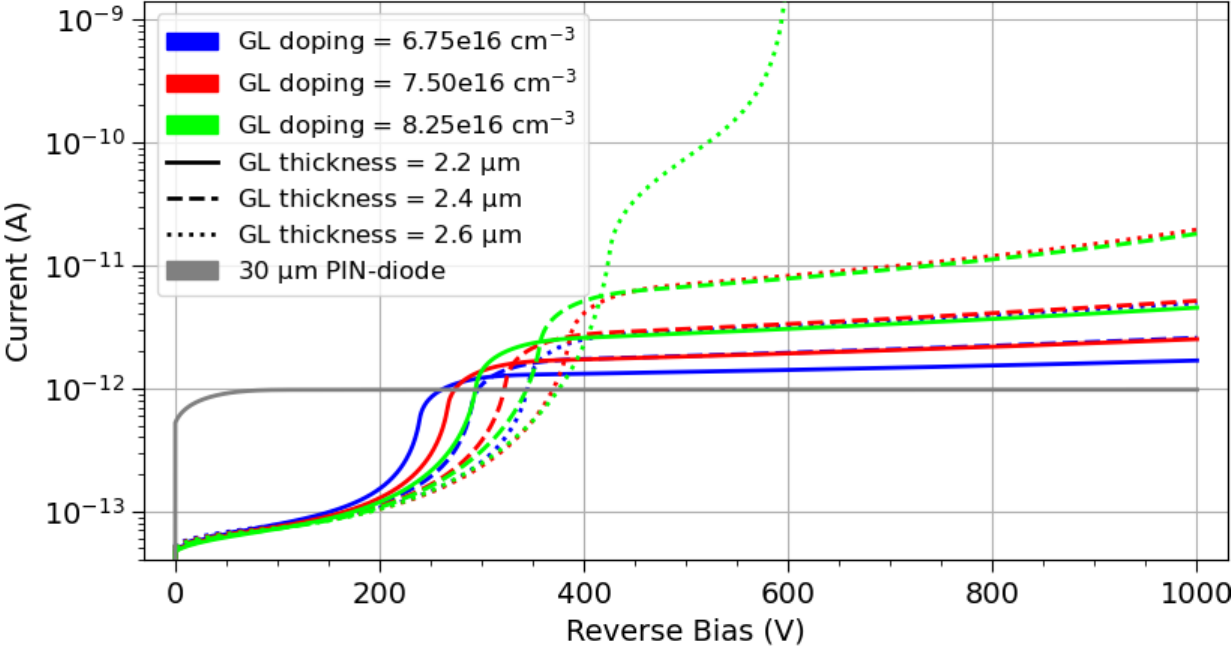}}
    \subfigure[C-V]{\includegraphics[width=0.7\linewidth]{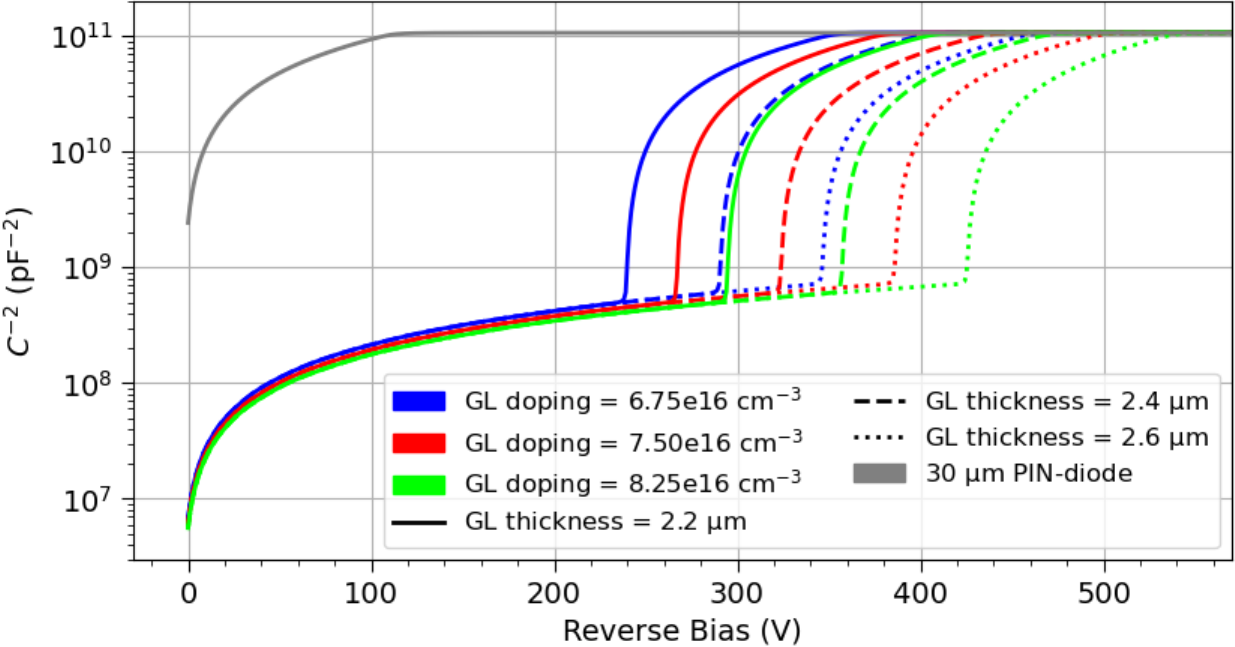}}
    \caption{Simulated I-V and $\mathrm{C^{-2}}$-V characteristics of a 4H-SiC LGAD without guarding structures compared to a $\SI{30}{\micro \meter}$ thick PIN diode.}
    \label{fig:I-V}
\end{figure}

Fig.\ref{fig:I-V} a and b show the simulated I-V and $\mathrm{C^{-2}}$-V characteristics of the proposed LGAD design in comparison to a $\SI{30}{\micro \meter}$ thick PIN diode without gain layer. Results are also shown for potential variations in the gain layer thickness ($\pm \SI{0.2}{\micro\meter}$) and doping concentration ($\pm \SI{10}{\percent}$) as reported by the manufacturer. A breakdown in the gain layer is only observed for the case with both maximum thickness and highest doping concentration. For all other configurations, the gain layer depletes at voltages below $\SI{400}{\volt}$, with full device depletion occurring below $\SI{500}{\volt}$, and dark current levels remaining under $\SI{30}{\pico \ampere}$.

\begin{figure}[ht!]
    \centering
    \includegraphics[width=0.7\linewidth]{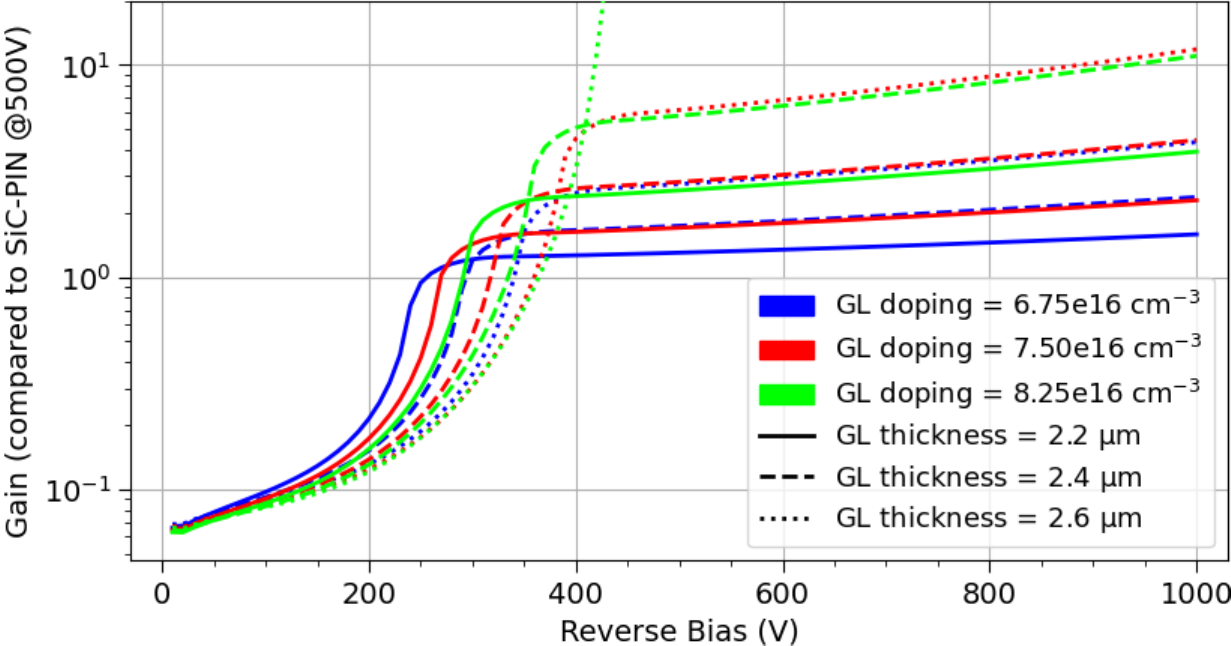}
    \caption{Simulated gain of a 4H-SiC LGAD without guarding structures, as a function of the applied reverse bias. Values are determined by comparing the respective integrated current pulse generated by a MIP in the LGAD structure to that of a $\SI{30}{\micro\meter}$ PIN diode at $\SI{500}{\volt}$ bias.}
    \label{fig:gain}
\end{figure}

Fig.\ref{fig:gain} depicts the simulated gain as a function of the applied reverse bias across the same range of gain layer variations. Excluding the single breakdown case identified above, the gain increase with reverse bias is rather smooth, while exhibiting a slightly steeper slope for thicker/more heavily doped gain layers, as one would expect. The simulations predict a signal gain in the range of approximately 1 to 10, indicating a stable amplification behavior within the expected parameter window.

\subsection{Breakdown Behavior}
\label{sec:breakdown_results}

\begin{figure}
    \centering
    \includegraphics[width=0.7\linewidth]{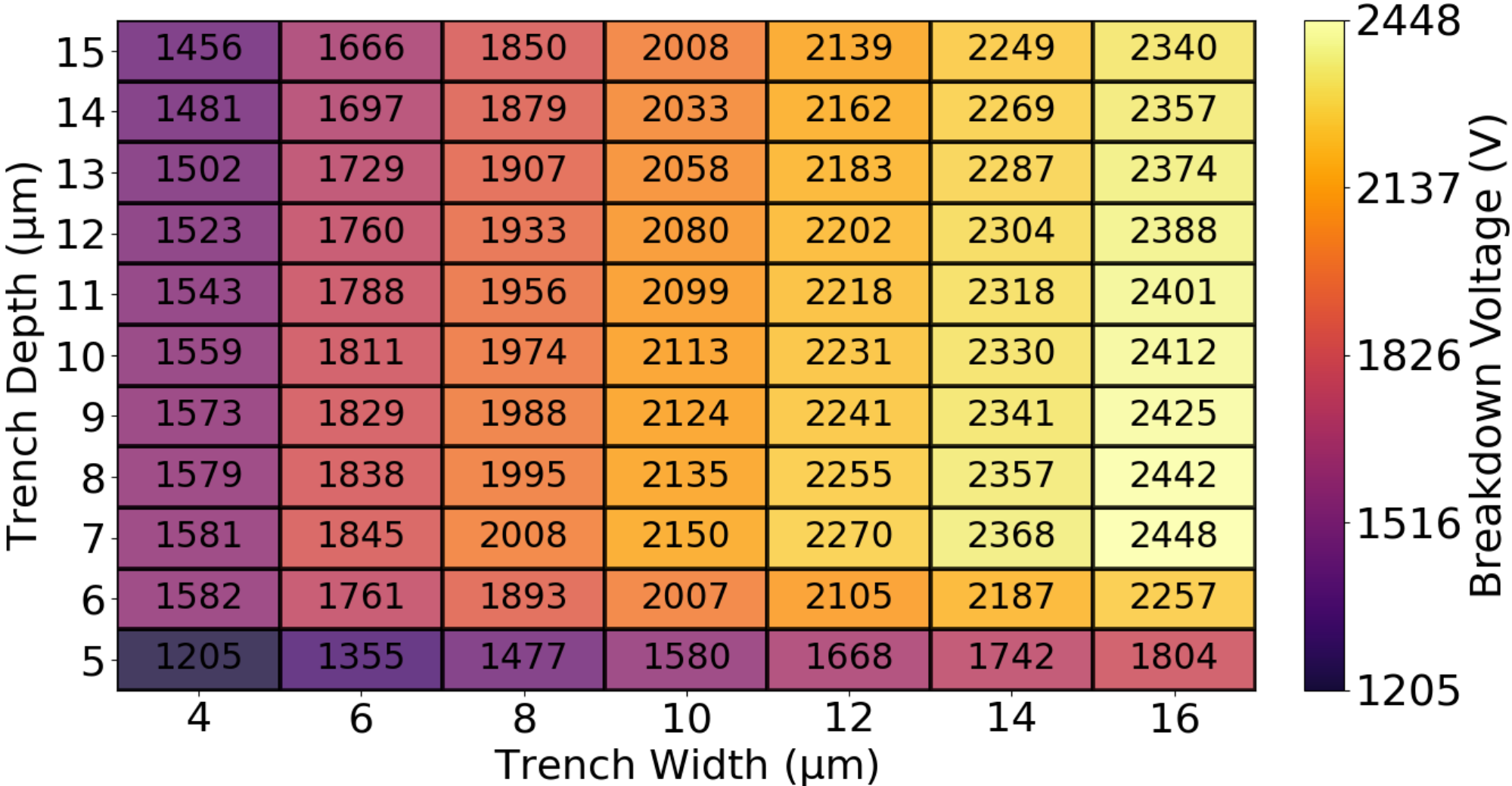}
    \caption{Breakdown voltage of the SiC LGAD design versus trench width and depth for a fixed $p+$-JTE implant with a width and depth of $\SI{30}{\micro\meter} \times \SI{4}{\micro \meter}$.}
    \label{fig:trench_map}
\end{figure}

Using the simulation geometry described in section \ref{sec:sim_w_guarding}, the breakdown behavior of the LGAD design was simulated for varying JTE implant widths, trench widths, and trench depths. A sweep over JTE widths revealed a clear saturation of the breakdown voltage beyond a width of $\SI{30}{\micro\meter}$, independent of the trench dimensions. This value was, therefore, selected as the JTE width for the final design. Fig.\ref{fig:trench_map} shows a heatmap of the simulated breakdown voltages for all considered combinations of trench widths and depths. While breakdown voltages above $\SI{1}{\kilo\volt}$ could be achieved across all configurations, the trench width showed a more pronounced impact on performance than the depth, which displayed a non-monotonic trend: The shallowest depth of $\SI{5}{\micro\meter}$ resulted in the earliest breakdown. Increasing the depth initially improved performance. But, for depths beyond $\SI{7}{\micro\meter}$, a gradual decline in breakdown voltage was observed. The optimal trench depth was therefore found to be $\SI{7}{\micro\meter}$. In contrast, the trench width consistently improved the breakdown voltage across the simulated range. The best performance was achieved at the largest trench width of $\SI{16}{\micro\meter}$. Combined with the optimal JTE width and trench depth, a maximum breakdown voltage of $\SI{2450}{\volt}$ could be obtained, demonstrating more than sufficient high-voltage stability. However, the final design, as given in Tab.\ref{tab:guard_design_parameters}, only features narrower variations of trenches. This is a result of restrictions by the manufacturer, as wider trenches inhibit larger risks of large-area surface damage during etching.

\section{Summary}
\label{sec:summary}

In this work, we presented the design and simulation process of a $\SI{30}{\micro\meter}$ thick 4H-SiC LGAD structure with trench isolation optimized for high-voltage stability. The $\SI{2.4}{\micro\meter}$ thick gain layer was realized through epitaxial growth rather than implantation, enabling a homogeneously doped profile that allows for full depletion below $\SI{500}{\volt}$. Key electrical characteristics, including I-V, C-V, and signal gain behavior, were simulated using a simplified quasi-1D device geometry, accounting for fabrication tolerances reported by the manufacturer.

To ensure robust edge termination, a guard structure employing $\SI{7}{\micro\meter}$ deep etched trenches was designed. For additional high-voltage stability, high-energy implanted $p^+$ JTEs with width and depth of $\SI{30}{\micro \meter} \times \SI{4}{\micro \meter}$ and a doping concentration of $\SI{1e17}{\centi \meter^{-3}}$ were integrated adjacent to the trenches. A parametric sweep over trench and implant dimensions identified the reported optimal configuration, achieving a suggested simulated breakdown voltage exceeding $\SI{2.4}{\kilo\volt}$. Fabrication of the final design is currently underway at IMB-CNM in Barcelona. 

\section*{Acknowledgements}

This work was supported by the Austrian Research Promotion Agency (FFG) in the project ``HiBPM'' (883652), by the Spanish Ministry of Science, Innovation and Universities (grant number CEX2023-001397-M), and by the European Union's ERDF program (MCIN/AEI/10.13039/501100011033/) ``A way of making Europe'' under grant reference PID2021-124660OB-C22.

\bibliographystyle{elsarticle-num}
\bibliography{bibliography} 

\begin{thebibliography}{10}
\expandafter\ifx\csname url\endcsname\relax
  \def\url#1{\texttt{#1}}\fi
\expandafter\ifx\csname urlprefix\endcsname\relax\def\urlprefix{URL }\fi
\expandafter\ifx\csname href\endcsname\relax
  \def\href#1#2{#2} \def\path#1{#1}\fi

\bibitem{nava2008}
F.~Nava, et~al., Silicon carbide and its use as a radiation detector material, Measurement Science and Technology 19~(10) (2008) 102001.
\newblock \href {https://doi.org/10.1088/0957-0233/19/10/102001} {\path{doi:10.1088/0957-0233/19/10/102001}}.

\bibitem{denapoli2022}
M.~De~Napoli, {SiC} detectors: {A} review on the use of silicon carbide as radiation detection material, Frontiers in Physics 10 (2022) 898833.
\newblock \href {https://doi.org/10.3389/fphy.2022.898833} {\path{doi:10.3389/fphy.2022.898833}}.

\bibitem{capan2022b}
I.~Capan, {4H}-{SiC} {Schottky} {Barrier} {Diodes} as {Radiation} {Detectors}: {A} {Review}, Electronics 11~(4) (2022) 532.
\newblock \href {https://doi.org/10.3390/electronics11040532} {\path{doi:10.3390/electronics11040532}}.

\bibitem{Gaggl2022}
P.~Gaggl, et~al., Charge collection efficiency study on neutron-irradiated planar silicon carbide diodes via {UV}-{TCT}, NIM-A 1040 (2022).
\newblock \href {https://doi.org/10.1016/j.nima.2022.167218} {\path{doi:10.1016/j.nima.2022.167218}}.

\bibitem{Gsponer_2023}
A.~Gsponer, et~al., {Neutron radiation induced effects in 4H-SiC PiN diodes}, JINST 18~(11) (nov 2023).
\newblock \href {https://doi.org/10.1088/1748-0221/18/11/C11027} {\path{doi:10.1088/1748-0221/18/11/C11027}}.

\bibitem{paper_HEPHY_0}
M.~Christanell, et~al., {4H-Silicon Carbide as particle detector for high-intensity ion beams}, JINST 17~(01) (2022).
\newblock \href {https://doi.org/10.1088/1748-0221/17/01/c01060} {\path{doi:10.1088/1748-0221/17/01/c01060}}.

\bibitem{Pellegrini_2014}
G.~Pellegrini, et~al., {Technology developments and first measurements of Low Gain Avalanche Detectors (LGAD) for high energy physics applications}, NIM-A 765 (2014) 12--16.
\newblock \href {https://doi.org/10.1016/j.nima.2014.06.008} {\path{doi:10.1016/j.nima.2014.06.008}}.

\bibitem{Zhao_2024}
S.~Zhao, et~al., {Electrical Properties and Gain Performance of 4H-SiC LGAD (SICAR)}, IEEE Transactions on Nuclear Science 71~(11) (2024) 2417--2421.
\newblock \href {https://doi.org/10.1109/TNS.2024.3471863} {\path{doi:10.1109/TNS.2024.3471863}}.

\bibitem{Yang_2025}
T.~Yang, et~al., {Ultra-Fast 4H-SiC LGAD with Etched Termination and Field Plate}, IEEE Electron Device Letters (2025) 1--1\href {https://doi.org/10.1109/LED.2025.3548509} {\path{doi:10.1109/LED.2025.3548509}}.

\bibitem{Novotny2025}
R.~Novotný, et~al., {First generation 4H-SiC LGAD production and its performance evaluation}, arXiv preprint (2025).
\newblock \href {https://doi.org/10.48550/arXiv.2503.07490} {\path{doi:10.48550/arXiv.2503.07490}}.

\bibitem{URL_CNM}
{IMB-CNM (Centro Nacional de Micr\'{o}electronica)-CSIC} (2024).
\newblock \href{http://www.cnm.es/}{[link]}.
\newline\urlprefix\url{http://www.cnm.es/}

\bibitem{URL_Sentaurus}
{Synopsys Sentaurus TCAD Ver. V-2023.09, Synopsys, Inc., Mountain View, CA, USA.}, \url{https://www.synopsys.com} (2025).

\bibitem{URL_mi2}
{mi2-factory GmbH, Jena, Germany}, \url{https://mi2-factory.com/} (2025).

\bibitem{URL_SiC_Sim}
P.~Gaggl, {Improving TCAD simulation of 4H silicon carbide particle detectors presented at 42nd RD50 workshop, Tivat, Montenegro}, \url{https://indico.cern.ch/event/1270076/contributions/5450202/ } (Jun 2023).

\bibitem{URL_LGAD}
P.~Gaggl, {TCAD simulations of 4H-SiC LGADs presented at 43rd RD50 workshop, CERN, Switzerland}, \url{https://indico.cern.ch/event/1334364/contributions/5672054/ } (Nov 2023).

\end{thebibliography}
\end{document}